\documentclass{aa}
\usepackage[varg]{txfonts}
\bibpunct{(}{)}{;}{a}{}{,} 
\usepackage{graphicx}	
\usepackage{amsmath}	
\usepackage{natbib}

\begin{document}

\title{\texttt{VBMicroLensing}: three algorithms for multiple lensing with contour integration}

\author{V. Bozza\inst{1,2}, V. Saggese\inst{2,3}, G. Covone\inst{2,3,4}, P. Rota\inst{1,2}, J. Zhang\inst{5}}

\offprints{V. Bozza, \email{valboz@sa.infn.it}}

\institute{Dipartimento di Fisica "E.R. Caianiello", Universit\`{a} di Salerno, Via Giovanni Paolo  132, Fisciano, I-84084, Italy 
\and Istituto Nazionale di Fisica Nucleare, Sezione di Napoli, Via Cintia, Napoli, I-80126, Italy
\and Dipartimento di Fisica "Ettore Pancini", Universit\`{a} di Napoli Federico II, Napoli, Italy
\and  INAF -- Osservatorio Astronomico di Capodimonte, Salita Moiariello 16, I-80131, Napoli, Italy
\and Department of Astronomy, Tsinghua University, Beijing 100084, China}

\date{Received/ Accepted}

\abstract {Modeling of microlensing events poses computational challenges for the resolution of the lens equation and the high dimensionality of the parameter space. In particular, numerical noise represents a severe limitation to fast and efficient calculations of microlensing by multiple systems, which are of particular interest in exoplanetary searches.}{We present a new public code built on our previous experience on binary lenses that introduces three new algorithms for the computation of magnification and astrometry in multiple microlensing.}{Besides the classical polynomial resolution, we introduce a multi-polynomial approach in which each root is calculated in a frame centered on the closest lens. In addition, we propose a new algorithm based on a modified Newton-Raphson method applied to the original lens equation without any numerical manipulation.}{These new algorithms are more accurate and robust compared to traditional single-polynomial approaches at a modest computational cost, opening the way to massive studies of multiple lenses. The new algorithms can be used in a complementary way to optimize efficiency and robustness.}{}

\titlerunning{\texttt{VBMicroLensing}: three algorithms for multiple lensing}
\maketitle 
  
\section{Introduction}\label{Sec intro}

Microlensing represents a unique possibility to investigate compact objects throughout our Galaxy and beyond by their gravitational effects regardless of their intrinsic luminosity \citep{Mao2012,Gaudi2012,Tsapras2018,Mroz2023,Covone2000}. It opens an avenue to the extension of  statistical studies of stellar populations, stellar remnants and planetary systems to smaller or darker objects that would be inaccessible by other methods. In the context of exoplanetary searches, microlensing stands out as the most effective method to detect planets at intermediate/wide orbits beyond the snow line \citep{Mroz2023}. Mass ratio functions derived by microlensing statistics \citep{Gould2010,Cassan2012,Suzuki2016,Suzuki2018} stand as the current benchmarks for testing planetary population synthesis models in the outer regions of planetary systems, where gaseous and icy giants form \citep{Mulders2019,Emsenhuber2021,Zhu2021}.

Within more than 200 microlensing planets found so far\footnote{See e.g. NASA exoplanet archive \url{https://exoplanetarchive.ipac.caltech.edu/}}, the number of evident multi-planetary events is remarkable, amounting to 7 (OGLE-2006-BLG-109 \citep{Gaudi2008,Bennett2010b}; OGLE-2012-BLG-0026 \citep{Han2013,Beaulieu2016}; OGLE-2014-BLG-1722 \citep{Suzuki2018b}; OGLE-2018-BLG-1011 \citep{Han2019}; OGLE-2019-BLG-0468 \citep{Han2022p}; KMT-2020-BLG-0414 \citep{Zang2021}; KMT-2021-BLG-1077 \citep{Han2022}). In addition, 11 planets were discovered in stellar binary systems \citep{2014ApJ...795...42P,2014Sci...345...46G,Bennett2016,2017AJ....154..223H,Bennett2020AJ....160...72B,2020AJ....159...48H,2021AJ....161..270H,2021AA...655A..24H,2022MNRAS.516.1704K,2024AA...685A..16H,2024arXiv240902157Z}. Summing the planets in these two categories, we get a sizeable fraction of 10.6\% planets in multiple systems, which forces us to take microlensing by multiple systems very seriously if we want to ensure that statistics of microlensing planetary systems are really complete. Moreover, with the advent of the {\it Roman} space telescope\footnote{\url{https://www.jpl.nasa.gov/missions/the-nancy-grace-roman-space-telescope}} providing exquisite photometry at 100ppm level, we can expect that many low-signal planets or far away companion stars or sometimes even moons will emerge and leave their signatures on a large fraction of planetary microlensing events. It is therefore urgent to develop fast and robust computational tools for multiple microlensing in order to be prepared to deal with such revolutionary data that would otherwise remain too problematic to be fully exploited.

State-of-art software in microlensing modeling and interpretation are based on a few packages that are now quite robust and well-tested for single and binary microlensing. Some of them are publicly available, such as adaptive contour integration by \citet{Dominik2007MNRAS.377.1679D} and \texttt{VBBinaryLensing}\footnote{\url{https://github.com/valboz/VBBinaryLensing}} by \citet{Bozza2}, also based on contour integration. For multiple lenses, inverse-ray shooting seems to be a more straightforward technique \citep{Kayser1986A&A...166...36K,Dong2006ApJ...642..842D,Gaudi2008,Han2013}. In fact, re-use of magnification maps generated with a large number of lenses can speed-up the calculations, but becomes very inefficient whenever orbital motion changes the configuration of lenses. In alternative, the image-centered ray-shooting method, put forward by \citet{BennettRhie1996,Bennett2010} has been successfully used in a great variety of microlensing events, including multiples \citep{Beaulieu2016,Bennett2016,Suzuki2018b,Bennett2020AJ....160...72B}. This algorithm, now also public\footnote{\url{https://github.com/golmschenk/eesunhong}}, relies on the resolution of the lens equation for the center of the source in order to find the centers of the images and optimize ray-shooting. However, when it comes to a system composed of $N$ lenses, the numerical resolution of the lens equation requires the extraction of all roots from a complex polynomial of degree $N^2+1$ \citep{Witt1990A&A...236..311W}. If lenses have very small mass ratios or very wide or short distances, numerical noise starts to become a real threat for 3 masses already. At this point, the only way to proceed seems to resort to quadruple precision floating point numbers, with a general slow-down of the code. In a similar way, a basic extension of contour integration from \texttt{VBBinarylensing} to triple lenses attempted by \citet{Kuang2021} suffers from similar numerical problems. In summary, at the present time, the analysis of multiple microlensing events remains extremely time consuming compared to binary microlensing, generally requiring heavy computational resources.

In this paper we present the three algorithms for multiple microlensing calculations that have been implemented in the new code \texttt{VBMicroLensing}\footnote{\url{https://github.com/valboz/VBMicroLensing}} that represents the evolution of \texttt{VBBinaryLensing} to multiple lenses. In Section 2 we review some basics about contour integration as performed by \texttt{VBBinaryLensing}. In Section 3 we also recall the Skowron \& Gould algorithm for the root extraction in polynomials.  In Section 4 we discuss our direct extension to multiple lenses with the traditional polynomial form of the lens equation. In Section 5 we present a new approach in which the polynomial is re-calculated in a frame centered on each lens in order to minimize numerical errors. In Section 6 we propose a new resolution paradigm on the original lens equation without any  reference to the polynomial form. In Section 7 we compare the performance and robustness of the different approaches. Then we conclude in Section 8.

\section{Contour integration}\label{Sec contour}

In a microlensing event, we may have photometric and astrometric measurements. Both require the detailed reconstruction of the images generated by gravitational lensing of the background source as caused by a foreground system of lenses. Assuming that all lenses can be modeled by point-like masses, the lens equation can be conveniently written in complex notation as \citep{Witt1990A&A...236..311W}
\begin{equation}
    \zeta = z -\sum_{i=1}^N\frac{ m_i}{\bar{z}-\bar{a}_i}, \label{LensEq}
\end{equation}
where the $N$ lenses have masses $m_i$ and angular coordinates $a_i$. All masses are in units of some reference mass $M$ (e.g. the total mass of the system); all angles are in units of the Einstein angle for this reference mass $M$, with their real and imaginary parts representing the two coordinates on the sky. The bar is used for complex conjugation. The position of the source is given by $\zeta$, while the images are found by solving Eq. (\ref{LensEq}) for $z$.

If we have a uniform brightness source with angular radius $\rho$, the flux observed by the observed will be magnified by the factor
\begin{equation}
\mu = \frac{A}{\pi \rho^2},
\end{equation}
where $A$ is the total area covered by the images.

\begin{figure}[h]
    \centering
    \includegraphics[width=0.4\textwidth]{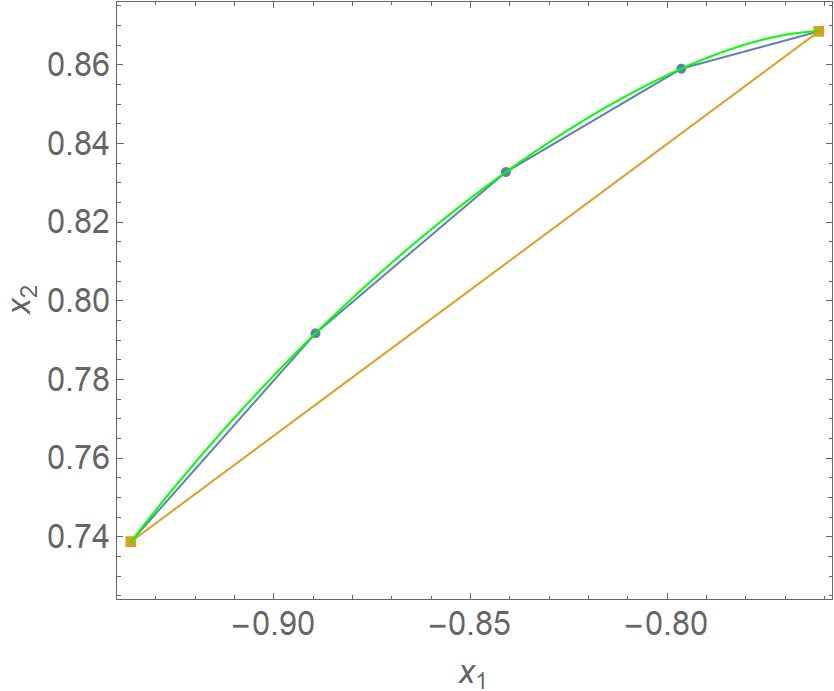}
    \caption{Illustration of the power of parabolic correction. Here we have an arc of an image in green. The straight line in orange would provide a poor approximation to the integral of the arc if we were limited to the trapezium approximation $A_T$. Adding the parabolic correction $A_P$, we improve the accuracy to the level that would be achieved by the trapezium approximation only with a four times denser sampling of the image contour (blue spline).  Therefore, we have much accurate estimates with fewer points (and shorter computational time).}
    \label{fig:parabolic}
\end{figure}

Contour integration calculates this area using Green's theorem \citep{Dominik1995,Dominik1998,GouldGaucherel}
\begin{equation}
    A= \frac{1}{2} \sum_I p_I \int_I z_I \wedge dz_I, \label{Green}
\end{equation}
where the sum covers all image contours labeled by the index $I$ with parity $p_I$. The wedge product of two complex numbers $a$ and $b$ is defined as 
\begin{equation}
    a \wedge b = \mathrm{Re}[a]\mathrm{Im}[b] - \mathrm{Im}[a]\mathrm{Re}[b] = \mathrm{Im}[\bar{a}b].
\end{equation}

The contour integral can be approximated if we start from a discrete sampling of the source boundary $\zeta_i = \zeta_c + \rho e^{i \theta_i}$, with $0=\theta_0<\theta_1< ... < \theta_n=2\pi$. For each point in this boundary we should solve the lens equation (\ref{LensEq}) as discussed in Section \ref{Sec SkowronGould} and get the corresponding images $z_{i,I}$. Then, the integral (\ref{Green}) is approximated by 
\begin{equation}
    A\simeq A_T +A_P,
\end{equation}
where the trapezium approximation
\begin{equation}
    A_T = \frac{1}{2} \sum_I p_I \sum_{I,i} z_{I,i} \wedge z_{I,i+1} \label{Trapezium}
\end{equation}
leaves a residual of third order in $\Delta\theta$. This is accounted by the parabolic correction \citep{Bozza1}
\begin{equation}
    A_{P_1} = \frac{1}{24} \sum_{I,i} \left[ \left( z'_{I,i} \wedge z''_{I,i} \right) + \left( z'_{I,i+1} \wedge z''_{I,i+1} \right) \right] \Delta\theta^3,
\end{equation}
which leaves a residual of fifth order in $\Delta\theta$ (see Fig. \ref{fig:parabolic} for an illustration of the benefits of parabolic correction).

Actually, in later versions of \texttt{VBBinaryLensing}, an alternative parabolic correction has been considered:
\begin{equation}
    A_{P_2} = \frac{1}{12} \sum_{I,i} ( z_{I,i+1}-z_{I,i})\wedge ( z'_{I,i+1}-z'_{I,i}) \Delta\theta.
\end{equation}

The two parabolic corrections $A_{P_1}$ and $A_{P_2}$ are generally equivalent, but in exceptional cases the first one vanishes when it should not (in particular at inflection points in the images). Therefore, we are now adopting the mean of the two forms $A_P=(A_{P_1} + A_{P_2})/2$ as a safer correction, while their difference is taken as an error estimator for each arc between $\theta_i$ and $\theta_{i+1}$.

Indeed, the main power of \texttt{VBBinaryLensing} resides in its reliable error estimators that are used to optimize the sampling of the source where it is most needed (e.g. close to caustics) without wasting time in oversampling everywhere.

Limb darkened sources are modeled by repeating the integration on concentric annuli \citep{Bozza1}. Additional error estimators drive the choice of the radii of these annuli.

If the source is far enough from the caustic, the difference between a point-source and a finite-source may fall below the tolerance set by the user. \texttt{VBBinaryLensing} also introduced a quadrupole test and a test on the ghost images (the spurious solutions of the polynomial, as explained in Section \ref{Sec SkowronGould}),  to check whether a finite-source calculation is needed \citep{Bozza2}. Most of the points in a light curve do not require a finite-source treatment and are thus calculated very rapidly.

Many other features have been introduced in the software over time, including astrometry, orbital motion for binary lenses and orbital motion for binary sources (xallarap) \citep{Bozza3}. 

\section{Resolution of the lens equation}\label{Sec SkowronGould}

Besides all the subtleties related to the finite-source treatment in the approximation of the contour integral (\ref{Green}), the core of the calculation remains the resolution of the lens equation (\ref{LensEq}), which has to be repeated for each point $\zeta_i$ in the source boundary sampling.

The standard way to proceed is to eliminate $\bar{z}$ using the complex conjugate of Eq. (\ref{LensEq}) and thus obtain a polynomial of degree $N^2+1$ \citep{Witt1990A&A...236..311W} (see Section \ref{Sec Singlepoly} for more details). Only the roots of this polynomial satisfying the original lens equation (\ref{LensEq}) are identified as physical images, whereas those roots that do not solve the original equation are just spurious solutions, sometimes called ghost images.

In a binary system of lenses $N=2$, the fifth degree polynomial can be written down in few steps. The roots of the polynomial can be found by standard Laguerre algorithm as described in the Numerical Recipes book by \citet{Press2002nrca.book.....P}. An alternative algorithm dynamically switching between Laguerre and Newton was proposed by \citet{SkowronGould} and became very popular in the microlensing community, being also adopted by \texttt{VBBinaryLensing} \citep{Bozza2}.

Whatever the chosen algorithm to solve the fifth degree polynomial, the accuracy of the roots rapidly degrades if one of the two lenses has a very small mass $m_2 \ll m_1$, which is just the planetary limit in which microlensing is mostly interested in. In fact, planetary images are formed very close to the planet $z \simeq a_2 - m_2/\left[\zeta-a_2+m_1/(a_1-a_2) \right]$. The summation of big and small terms is exposed to numerical truncation, leading to inaccurate images positions. For a binary lens, the simple turnaround is to set the origin of the reference frame in the planet $a_2=0$. In this way, even in extreme planetary regimes with $m_2/m_1 \ll 1$, the planetary images forming close to the second mass are recovered with very high accuracy.

There is also some interest in alternative algorithms, such as the Aberth-Ehrlich root finding scheme, which could be a promising avenue to speed up the calculations keeping high accuracy  \citep{Fatheddin2022MNRAS.514.4379F}, but a fair comparison is yet to be done.

\section{Algorithms for multiple lenses: \texttt{Singlepoly}}\label{Sec Singlepoly}

With little algebraic manipulation, we can write down the general expression for the polynomial equivalent to the multiple lens equation (\ref{LensEq}):
\begin{equation}
    (\zeta - z) \prod_i h_i(z) + \sum_i m_i \prod_{l \neq i} h_l(z)\prod_j (z-a_j) =0, \label{polynomial}
\end{equation}
with
\begin{equation}
    h_i(z)=(\bar \zeta - \bar a_i) \prod_j (z-a_j) + \sum_j\left[m_j\prod_{k\neq j}(z-a_k) \right].
\end{equation}
Since $h_i(z)$ is a polynomial of degree $N$ in $z$, the polynomial in Eq. (\ref{polynomial}) is of degree $N^2+1$, as anticipated \citep{Witt1990A&A...236..311W}.

An immediate generalization of the algorithms in \texttt{VBBinaryLensing} is indeed possible. The roots of the polynomial (\ref{polynomial}) can be found by the same \citet{SkowronGould} algorithm. We already pointed that the numerical accuracy is rapidly degraded even for one planet if the reference frame is not centered on the planet. Therefore, we can still move the reference frame in the smallest mass and get accurate roots around it. However, if there are two or more planets in the system, this algorithm is bound to fail on the roots closer to the other planets. This is the main motivation for the alternative algorithms presented in the two following sections, which are designed so as to solve this problem. Nevertheless, for systems composed by similar masses and at most one planet, the roots of the polynomial centered on the lowest mass are sufficiently accurate to represent a benchmark for other algorithms. Therefore, we decided to include the single-polynomial algorithm in \texttt{VBMicroLensing} as a useful reference, although it will be rarely employed in any scientific analysis.

The code based on the resolution of the polynomial (\ref{polynomial}) can be selected by the user and is labeled as \texttt{Singlepoly}. The strategy for finite-source calculations is exactly the same as in \texttt{VBBinaryLensing}. We adopt parabolic corrections and error estimators calculated in complete analogy with the binary lensing code. Therefore, the \texttt{Singlepoly} code benefits from an optimized sampling and accuracy control. The only difference to be mentioned is in the error estimator described in Section 3.3 of the paper by \citet{Bozza1}, introduced to catch temporary images for sources almost tangent to fold caustics. This estimator was based on a distance check on the two ghost images before caustic crossing. In the context of multiple lensing, in general we have many ghost images (from a minimum of $N^2-5N+4$ to a maximum of $N^2-N$). So we select the pair with the minimum mutual distance and use them in this error estimator.

As mentioned in the introduction, \citet{Kuang2021} have developed a triple-lens code for microlensing using contour integration and based on the \citet{SkowronGould} algorithm for the resolution of the polynomial. There are several differences between their code and our \texttt{Singlepoly} algorithm in \texttt{VBMicroLensing}: they do not implement any parabolic correction, thus requiring denser sampling to achieve the same accuracy; they have no error estimators, thus relying on uniform sampling and a check that two consecutive runs return similar results within the tolerance; they have a different image reconnection algorithm. In general, a less efficient sampling and the lack of parabolic correction would lead to longer computations to reach similar accuracy. This is indeed verified in Section \ref{Sec performance}, which includes a comparison with the code \texttt{triplelens} by \citet{Kuang2021}. 

\begin{figure*}[t]
    \centering
    \includegraphics[width=0.7\textwidth]{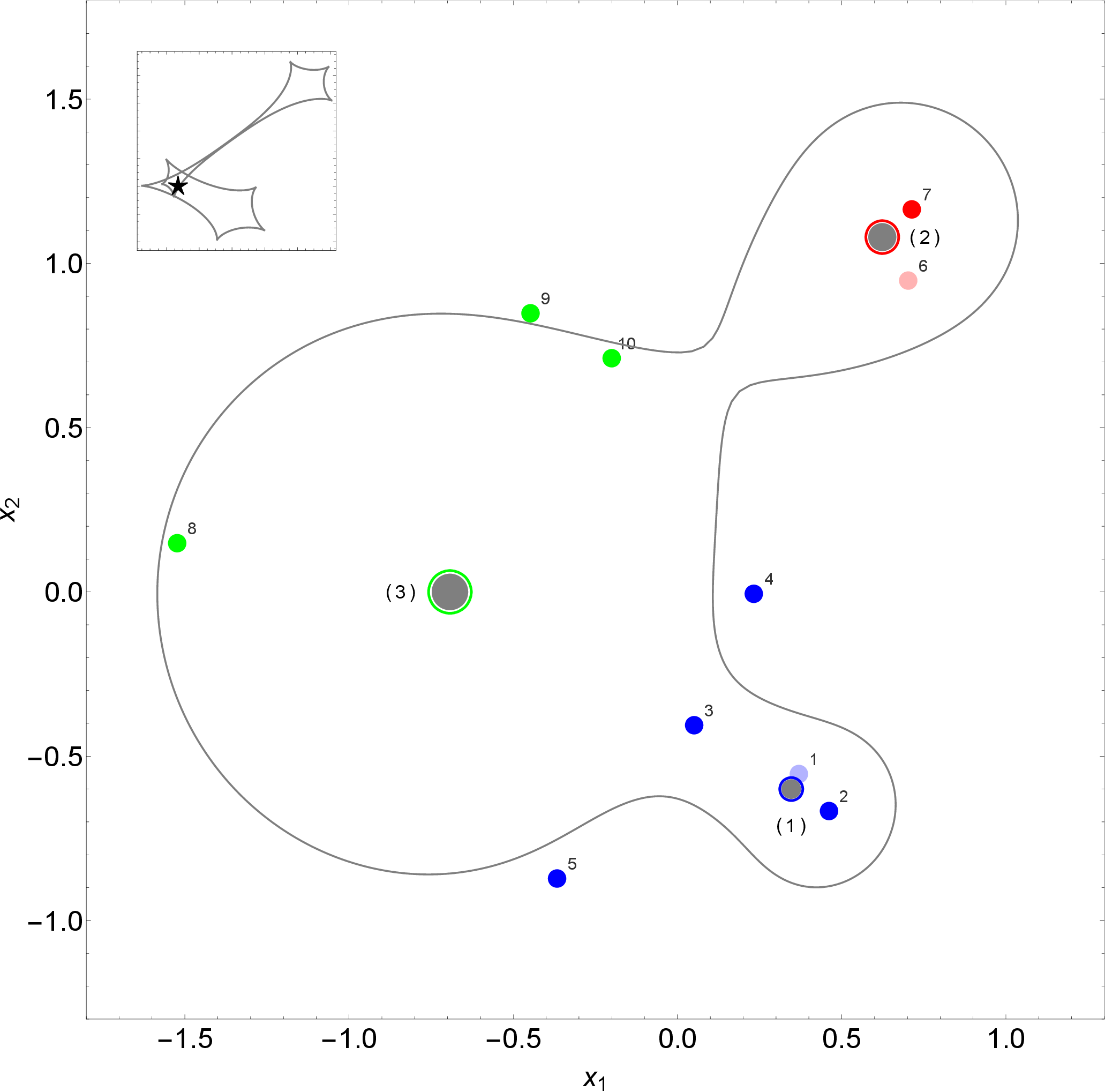}
    \caption{The \texttt{Multipoly} algorithm performs a roots search by exploring a configuration of three lenses. These lenses are marked in gray, numbered (1), (2), (3) in order of increasing mass, reflecting the sequential order in which the algorithm constructs the polynomials centered on each respective lens. The respective critical curve is also shown. The source position is indicated in the caustics box. The algorithm begins its search for roots near the smallest lens and gradually moves towards the more massive lenses. The order in which the roots are found is denoted by the numbers, while the color corresponds to the polynomial from which they were derived; the polynomial that is centered in the lens with a border of the same color. Opaque-colored roots represent real images, while transparent roots are classified as ghost images.}
    \label{fig:Multipoly}
\end{figure*}

\section{Algorithms for multiple lenses: \texttt{Multipoly}}\label{Sec Multipoly}

The \texttt{Multipoly} algorithm represents an advancement over the classical polynomial method \texttt{Singlepoly}. It addresses the inherent limitations of solving the lens equation (\ref{LensEq}) by dynamically re-centering the polynomial computation on each lens within the multiple system. This method aims to enhance accuracy by isolating the influence of individual lenses, minimizing numerical errors that arise in complex multi-lens configurations.
Rather than solving a single global polynomial, the \texttt{Multipoly} algorithm recalculates the polynomial in local coordinate systems centered sequentially on each lens. This local re-centering approach, even in planetary regimes with $m_i/m_j \ll 1$, enhances the accuracy of the polynomial's root calculations, particularly in regions near the reference center.

The \texttt{Multipoly} method begins by selecting the lens with the smallest mass as the origin (center) of the local coordinate system. It then constructs and solves a polynomial equation specific to this local frame, focusing on roots that are relevant in the vicinity of this lens. When a root is found to be closer to another lens, it iteratively transitions to the next lens with increasing mass. This process continues until the final, highest-mass lens is reached. For this last lens, the polynomial is divided by the previously found roots to determine any remaining roots, ensuring all potential images in the multi-lens system are accurately accounted for. This iterative approach to polynomial computation is detailed in Fig. \ref{fig:Multipoly}.

Compared to the \texttt{Singlepoly} approach, the \texttt{Multipoly} algorithm prioritizes enhanced accuracy over computational speed (see Section \ref{Sec performance}). While the \texttt{Singlepoly} method remains efficient for simpler, less complex lens configurations, the \texttt{Multipoly} algorithm excels in scenarios that require higher levels of accuracy, particularly when dealing with multi-lens systems that have small lens separations or significant mass ratios between lenses. By focusing the polynomial computation on regions near each lens, the \texttt{Multipoly} method can more accurately capture the complexities introduced by multiple lenses.

The \texttt{Multipoly} method incorporates robust error estimation techniques similar to those in the \texttt{Singlepoly} and offers the advantage, unlike the \texttt{Nopoly} method described in Section \ref{Sec Nopoly}, of preserving information about ghost images generated during the computation process. This feature enables the use of error estimators based on ghost images. A more detailed comparison of the performance of \texttt{Multipoly} and other algorithms is provided in Section \ref{Sec performance}.

\section{Algorithms for multiple lenses: \texttt{Nopoly}}\label{Sec Nopoly}

The resolution of the multiple lens equation (\ref{LensEq}) has been always attempted either by inverse-ray-shooting or by transforming it to the polynomial (\ref{polynomial}) to be solved by some root finding algorithm. As we have realized with the two algorithms presented up to now, the transformation to the polynomial requires long computations in which the numerical accuracy is gradually degraded. In addition, algorithms based on extraction of roots and consecutive division of the polynomial to find the following root, further degrade numerical accuracy. The \texttt{Multipoly} algorithm ensures that numerical accuracy is met for each root at the cost of re-calculating the polynomial for each lens, resulting in longer computations in the name of robustness. The fact that the number of roots scales quadratically with the number of lenses $N$, while the number of images only scales linearly provides further motivation to look for alternative approaches that eliminate the transformation to the polynomial.

\subsection{A direct Newton-Raphson approach}\label{Sec NR}

The algorithm \texttt{Nopoly} that we are describing in this section starts from the consideration that a Newton-Raphson method can be directly applied to the original lens equation (\ref{LensEq}) without any transformations. In order to shorten expressions, we make the following definitions:
\begin{equation}
S_q(z) \equiv  \sum_{i=1}^N\frac{ m_i}{\left(z-a_i\right)^q},  \label{Sq}    
\end{equation}
\begin{equation}
L(z) \equiv  \bar{\zeta}-\bar{z}+S_1(z), \label{Lz}
\end{equation}
\begin{equation}
J(z) \equiv  1-S_2(z)\bar{S}_2(z). \label{JS2}
\end{equation}

$L(z)$ will vanish if $z$ solves the lens equation, i.e. it is an image of the source $\zeta$. $J(z)$ is the Jacobian determinant of the lens equation, which vanishes at critical curves.

Suppose that we start from some initial condition $z_{k}$ and we want to move by $\epsilon$ in such a way that $z_k+\epsilon$ solves the lens equation (\ref{LensEq}). Then, expanding to first order in $\epsilon$, and using the definitions (\ref{Sq})-(\ref{JS2}), we find the equation
\begin{equation}
    \bar{L}(z_k) = \epsilon + \bar{S}_2(z_k)\bar{\epsilon}.
\end{equation}

Coupling with the complex conjugate equation, we can solve for $\epsilon$ and $\bar{\epsilon}$ to find
\begin{equation}
    \epsilon = \frac{\bar{L}(z_k)-\bar{S}_2(z_k)L(z_k)}{J(z_k)}. \label{epsilon}
\end{equation}

Therefore, if $z_k$ is close enough to a solution of Eq. (\ref{LensEq}), the complex number
\begin{equation}
    z_a\equiv z_k + \epsilon
\end{equation}
should represent a better approximation to the solution. By iterating this process, in principle, if we start from a generic initial condition we should find the closest image with an arbitrary accuracy. The appeal of this algorithm is that no manipulations of the original lens equation are needed, thus preserving numerical accuracy to the highest possible level. Indeed, if we use this algorithm to polish roots previously found by the polynomial approach, we can already experience a great benefit. Yet, we are tempted to move beyond and exclude the polynomial approach at all by building a new framework with this Newton-Raphson approach as the core algorithm.

\begin{figure}[t]
    \centering
    \includegraphics[width=0.45\textwidth]{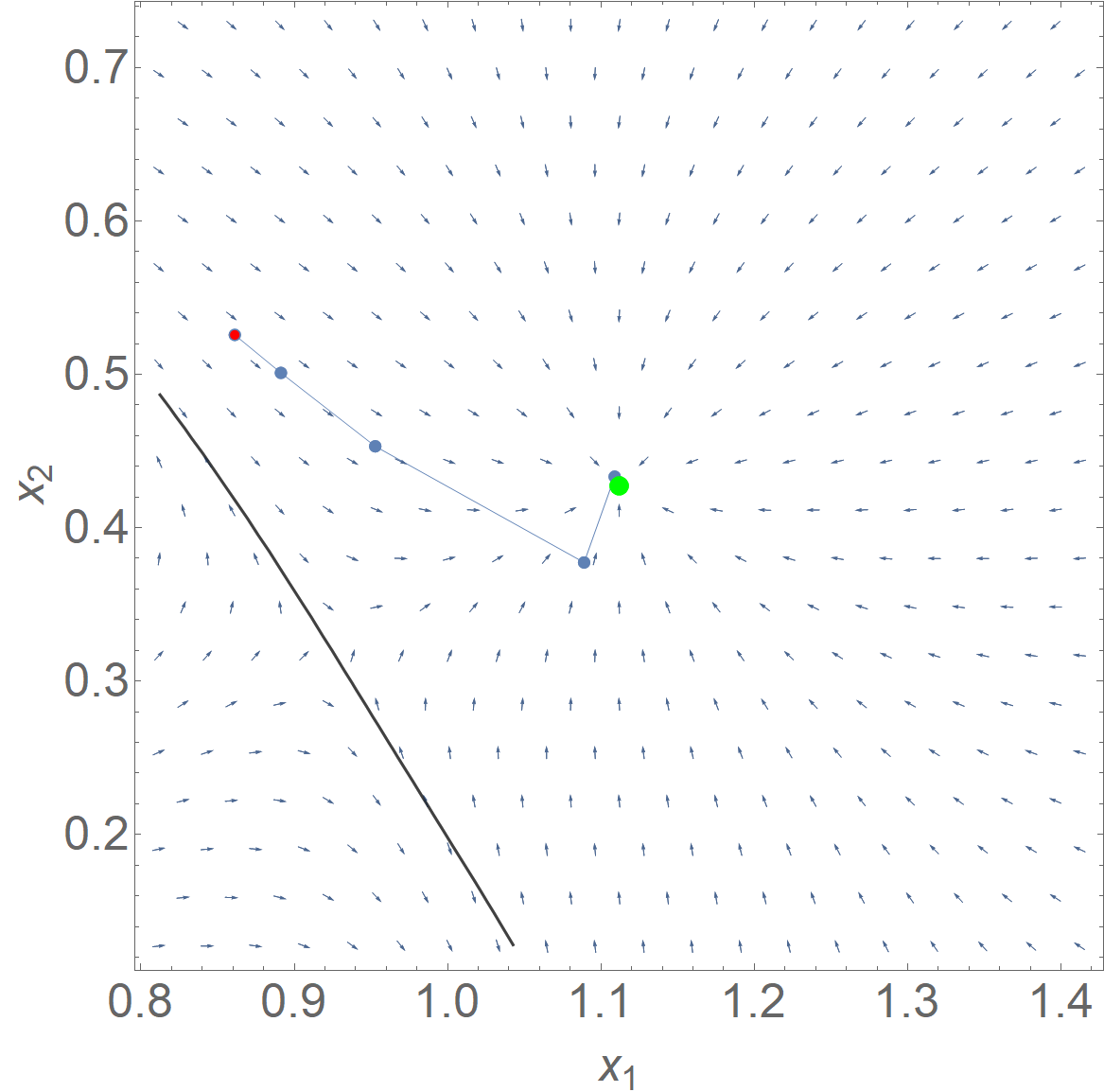}
    \caption{Illustration of the Newton-Raphson method starting from an initial seed (red dot) and reaching a solution of the lens equation (green dot) in 5 steps. The vector field indicates the direction of $\epsilon$ as calculated by Eq. (\ref{epsilon}). Firstly, we note that the field locally converges to the solution. However, beyond a nearby critical curve (the gray line in the bottom left) the vector flow diverges from this solution. This is a good reason to forbid critical curve crossing (Section \ref{Sec critical curve}).}
    \label{fig:epsilon}
\end{figure}

Unfortunately, the lens equation is complicated enough to make this framework far from trivial. In order to find all images, we need to repeat the search from a large enough set of initial conditions. We will eventually need to remove duplicates and check that no images are missing. Furthermore, we have to introduce special corrections to avoid divergences and oscillating behaviors that would make the simple Newton-Raphson algorithm fail. Fig. \ref{fig:epsilon} illustrates a basic chain reaching a solution of the lens equation in few steps from an initial seed. 

\subsection{Corrections to the Newton-Raphson algotithm}\label{Sec corrections}

\subsubsection{Oscillations}\label{Sec oscillations}
Once we calculate the proposed step $\epsilon$ by Eq. (\ref{epsilon}), we check that the new point is not too close to the previous two points in the chain. If $|z_k+\epsilon - z_{k-1}|<|\epsilon|$, we decrease the step $\epsilon$ by a pseudorandom number taken from a fixed sequence.

\subsubsection{Avoid explosion}\label{Sec explosion}
Another requirement is that the new step $\epsilon$ cannot be more than twice longer than the previous step. This ensures that the chain will not diverge if the denominator of $\epsilon$ (note that this denominator is just the Jacobian determinant of the lens equation) is very small, which may happen if we get too close to a critical curve. 

Actually, this condition is relaxed if the current Jacobian $J_k$ is larger than the old Jacobian $J_{k-1}$, i.e. if we are moving away from a critical curve. In practice, we define a factor
\begin{equation}
    f=\frac{|\epsilon|}{|z_k-z_{k-1}|\left[1+\mathrm{min}(1,J_k/J_{k-1}) \right]}.
\end{equation}
If $f>1$, we divide $\epsilon$ by $f$.

\subsubsection{Avoid critical curve crossing}\label{Sec critical curve}

For a more efficient search, in order to make sure that all different regions inside and outside critical curves are fully explored, we wish to keep each individual chain within the same region. Therefore, we forbid critical curve crossing.

If the sign of $J(z_k+\epsilon)$ is different from the sign of $J(z_k)$, we calculate a correction $\delta$ such that $z_k+ \epsilon+\delta$ is back in the original Jacobian region. In order to do that, we need an approximated distance to the closest critical curve. In the case of a single mass, the calculation is quite trivial. Suppose we are at some point $z$, the Jacobian would simply be
\begin{equation}
    J=1-\frac{m_1^2}{z^2\bar{z}^2}.\label{J1}
\end{equation}
We know that the critical curve is a circle with radius $\sqrt{m}$, so the closest zero of the Jacobian to $z$ would be at
\begin{equation}
    z_0=\sqrt{m_1}\frac{z}{\sqrt{z \bar{z}}}. \label{z0J1}
\end{equation}
Of course, by construction, we have $J_1(z_0)=0$.

Now we want to express this point $z_0$ using only local information in $z$. Indeed, we can use Eq. (\ref{J1}) to express the mass in terms of the Jacobian
\begin{equation}
    m_1=z\bar{z}\sqrt{1-J}.
\end{equation}
Substituting in Eq. (\ref{z0J1}), we have 
\begin{equation}
    z_0=z\left(1-J\right)^{1/4}. \label{z02J1}
\end{equation}
Then, we can also use the information that in our simple single-lens case 
\begin{equation}
    z=\frac{(m_1/z^2)}{m_1/z^3)} = S_2/S_3,
\end{equation}
with $S_q$ defined in Eq. (\ref{Sq}). Subtracting the starting point $z$, we have
\begin{equation}
    \delta_0 \equiv z_0-z = \frac{S_2}{S_3}\left[\left(1-J\right)^{1/4} -1 \right].\label{delta0}
\end{equation}
This expression is exact for a single-lens. However, in a multiple lens system, most critical curves will be dominated by one lens, or the combination of two close lenses. In both cases, the single-lens approximation would provide a very good approximation to the local distance to the closest critical curve. Even in those sections of critical curves where the contributions by two or more lenses is relevant, the order of magnitude of the distance will remain the same as this $\delta_0$. Note that for $J$ very close to zero, the expansion of Eq. (\ref{delta0}) would be
\begin{equation}
    \delta_0\simeq -J\frac{S_2}{S_3},
\end{equation}
which would be a linear extrapolation by the tangent method. The expression $\delta_0$ works much better also when the point is far from the critical curve and we have to calculate a correction that is just right without overshooting or undershooting.

Now, coming back to our problem, we want to push $z_k+\epsilon$ back in the same Jacobian region. Then we adopt
\begin{equation}
    \delta = \delta_0 \left[1-\frac{J(z_k)}{J(z_k+\epsilon)} \right], \label{delta}
\end{equation}
i.e. we try to push back the chain so that $z_k+\epsilon+\delta$ recovers the original value of the Jacobian of $z_k$.

\begin{figure}[t]
    \centering
    \includegraphics[width=0.45\textwidth]{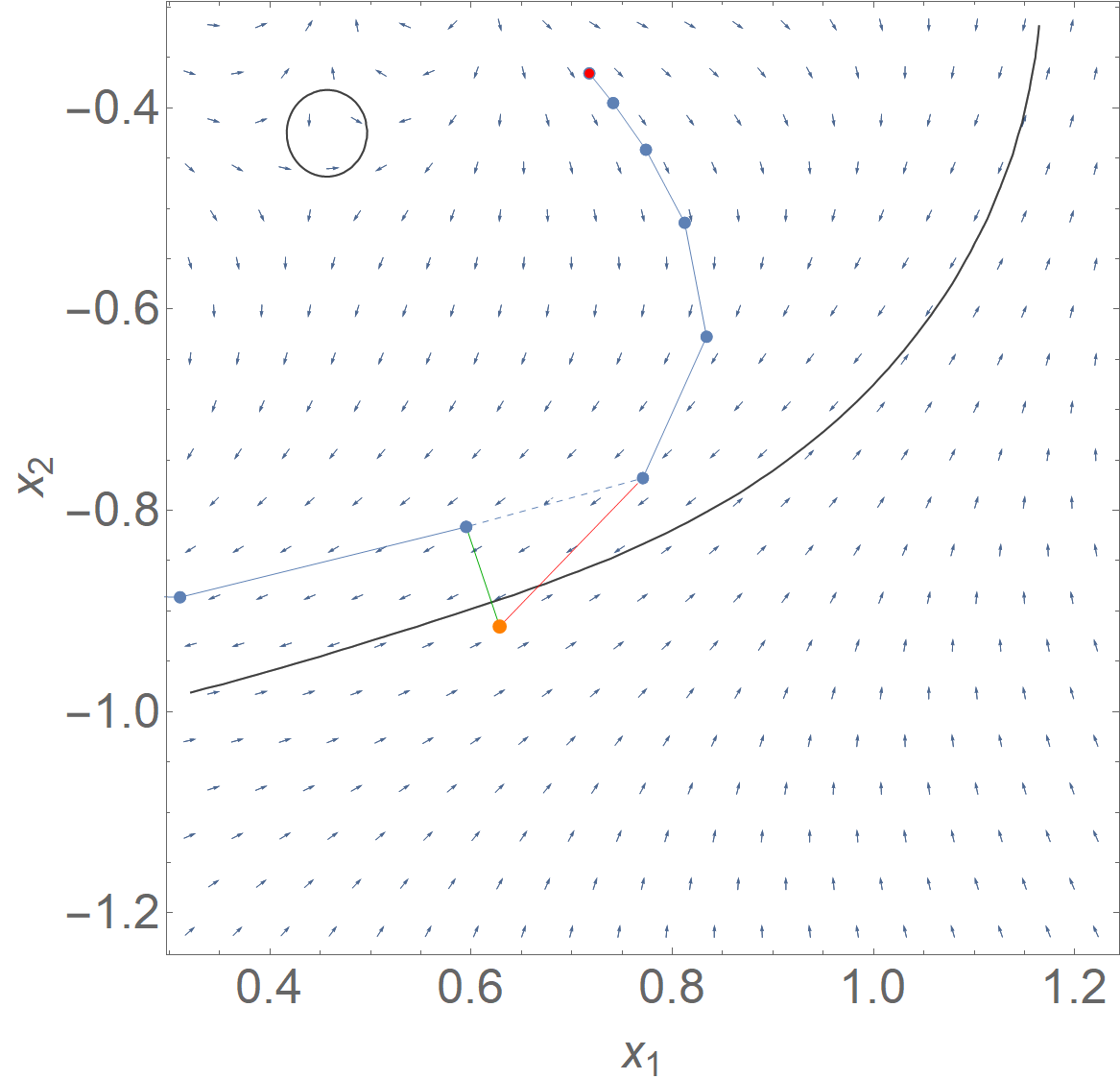}
    \caption{In this chain starting from the red initial seed, after 5 steps, the $\epsilon$ from Eq. (\ref{epsilon}), represented here by the red step, would cross the critical curve (the gray line). At this point, the $\delta$ correction (\ref{delta}), represented by the green step, pushes the chain back to the same Jacobian region, ensuring that the exploration continues without missing any solutions. }
    \label{fig:delta}
\end{figure}

Of course, we immediately check that the correction is successful. If the sign of the Jacobian remains opposite, we increase $\delta$. If the correction brings the new point too close to $z_k$, we shorten $\delta$ to avoid oscillating behavior. Fig. \ref{fig:delta} shows an example of critical curve crossing avoided by the $\delta$-correction.

\subsubsection{Further corrections if next point is worse than previous}

As in all step-by-step algorithms, it may happen that the new point $z_k+\epsilon+\delta$ does not improve the resolution of the lens equation, i.e. $|L(z_k+\epsilon+\delta)|$>$|L(z_k)|$. Here we distinguish two possibilities depending on the behavior of the Jacobian gradient
\begin{equation}
    dJ \equiv \frac{\partial J}{\partial z} = 2S_3 \bar{S}_2.
\end{equation}

If the direction of the gradient has changed, we have likely ended up in a very different region of the lens equation. This is not particularly good if we want to make sure that our exploration remains local around our initial condition. Therefore, if we find that the scalar product
\begin{equation}
    \mathrm{Re}[dJ(z_k)]\mathrm{Re}[dJ(z_k+\epsilon+\delta)]+\mathrm{Im}[dJ(z_k)]\mathrm{Im}[dJ(z_k+\epsilon+\delta)]<0
\end{equation}
then we halve $\epsilon$ until we get an improvement.

\begin{figure}[t]
    \centering
    \includegraphics[width=0.45\textwidth]{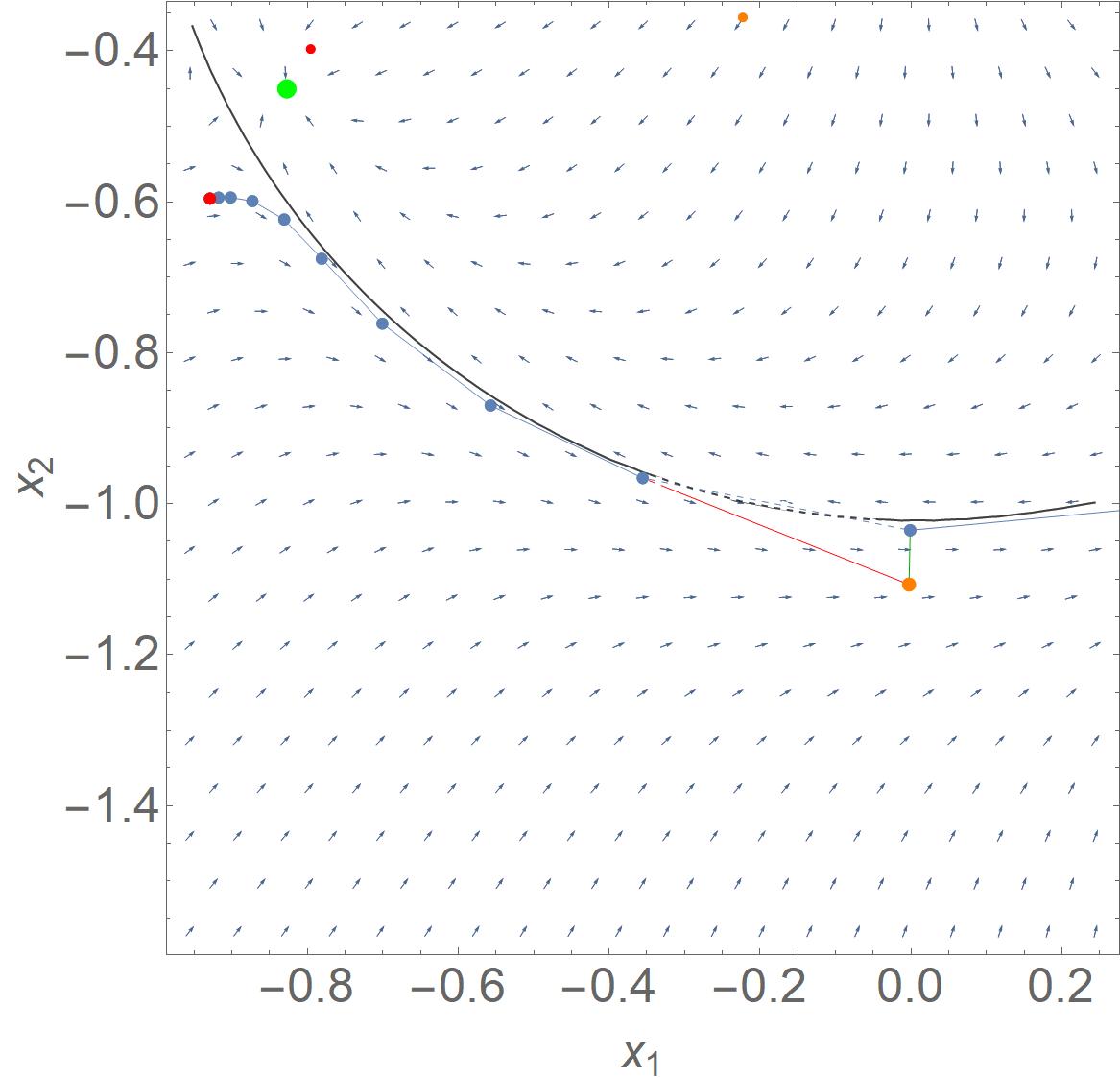}
    \caption{In this chain starting from the red initial seed, after a few steps grazing a critical curve, the $\epsilon$ from Eq. (\ref{epsilon}), represented here in orange is too long and worsens the fulfillment of the lens equation. At this point, the $\kappa$ correction (\ref{kappa}), represented by the green step, pushes the chain closer to the critical curve, where the lens equation is closer to be fulfilled. Such correction is important to accurately follow the vector flow all around the critical curves. }
    \label{fig:kappa}
\end{figure}

On the other hand, if we are in the same gradient region, this occurrence can be simply due to the fact that the critical curve has a curvature, while our steps are rectilinear. Instead of shortening the step, here we just push the point closer to the critical curve, by an amount dictated by the lens equation. In practice, defining the gradient of the squared modulus of the lens equation
\begin{equation}
    dL_2 \equiv \frac{\partial |L|^2}{\partial z} = -\bar{L} + \bar{S}_2 L,
\end{equation}
our correction is 
\begin{equation}
    \kappa = - \frac{|L|^2}{\mathrm{Re}[dJ]\mathrm{Re}[dL_2]+\mathrm{Im}[dJ]\mathrm{Im}[dL_2]} dJ \label{kappa}
\end{equation}
As usual, we halve $\kappa$ if it takes the point beyond the critical curve. Fig. \ref{fig:kappa} shows a situation in which we stick closer to the critical curve thanks to the $\kappa$-correction.

In the end, for most steps, we will simply set $z_{k+1}=z_k+\epsilon$ with $\epsilon$ determined by Newton-Raphson algorithm (\ref{epsilon}), but we may have steps where one among $\delta$ (\ref{delta}) and $\kappa$ (\ref{kappa}) corrections is needed or even both and then we have to set $z_{k+1}=z_k+\epsilon+\delta+ \kappa$.

\subsection{Stopping criteria}

The search for the solution is stopped when one of the following conditions is met:
\begin{itemize}
    \item the Newton-Raphson displacement (\ref{epsilon}) is smaller than an accuracy threshold fixed at $|\epsilon|<3\times10^{-11}$;
    \item the corrected displacement after checks for oscillations and explosion (Sects. \ref{Sec oscillations} and \ref{Sec explosion})  is comparable to machine precision $|\epsilon|<10^{-14}$;
    \item the lens equation is already satisfied to machine precision $|L(z_k)|<10^{-15}$;
    \item the number of consecutive corrections to $\delta$ to avoid oscillations (see end of Sect. \ref{Sec critical curve}) exceeds 9;
    \item the total number of iterations exceeds 100.
\end{itemize}

If one of the first three criteria is fulfilled, the solution is considered as acceptable and compared to the previously found solutions (Sect. \ref{Sec duplicates}). If one of the last two criteria is fulfilled, the current solution is rejected since convergence has not been achieved.

\subsection{Error assessment}

Each solution is accompanied by an uncertainty given by 
\begin{equation}
    Err= |\epsilon| + \frac{3\times10^{-15}}{|J|}. \label{Err}
\end{equation}

The first term is the size of the last Newton-Raphson step (\ref{epsilon}), while the second comes from the fact that if the terms in the lens equation are known to machine precision, $\epsilon$ cannot be calculated better than the machine precision divided by the Jacobian.

\subsection{Removal of duplicates}\label{Sec duplicates}

Each solution found is immediately compared to previously found solutions to check for duplicates. If $z$ is the freshly found solution and $z_i$ are the previously found ones, we check that for each solution with the same parity as $z$ we have
\begin{equation}
    \frac{|z-z_i|^2}{Err^2+Err_i^2 + 4\times 10^{-20}}>10.
\end{equation}

If this condition is satisfied for all previously found solutions with the same parity, then $z$ is considered as a new independent solution. The numerical constant in the denominator matches our original requirement to find solutions of the lens equation with an approximation of the order $10^{-10}$. The threshold on the right hand side removes any ambiguous solutions.

If one of the previous solutions $z_{\bar{i}}$ falls below the threshold, then the new solution $z$ is a duplicate of $z_{\bar{i}}$. We will only retain the solution with the smaller error, as estimated by Eq. (\ref{Err}).

\subsection{Initial conditions}

Once we have set up an effective and robust algorithm that converges to an image from a given initial condition, we now need to ensure that all images are actually found. In principle, we can solve this problem by starting from a huge number of initial seeds. However, if we do not want to spoil the efficiency of the algorithm, we need to make a clever choice guided by the mathematics of the lens equation. Here we present our initial conditions divided into four consecutive levels of searches.

\subsubsection{Seeds for basic images} \label{Sec basic images}

The minimum set of images for a multiple lens equation with $N$ lenses when the source is outside any caustics is given by one positive parity image and $N$ negative parity images, one for each lens. Therefore, the first level search starts from points that try to approximate such a configuration.

The guess for the positive parity image is
\begin{equation}
    z_+ \equiv \zeta + \sum_i \frac{\zeta- a_i}{2} \left[\sqrt{1+\frac{4m_i}{|\zeta- a_i|^2}}-1 \right].
\end{equation}
For a single-lens this is the exact position of the positive parity image. For more lenses it can be used as a first guess by summing the displacement caused by all lenses. If this guess does not lie in a region with $J>0$, we displace it in the $\zeta$ direction until $J(z_+)>0$.

For the negative parity images we use
\begin{equation}
    z_{-i} \equiv a_i -  \frac{\zeta- a_i}{2} \left[\sqrt{1+\frac{4m_i}{|\zeta- a_i|^2}}-1 \right],\label{z-i}
\end{equation}
i.e. the negative parity image of the single-lens equation for each lens. If this guess is not in a region with $J<0$, we halve the displacement from the lens position $a_i$ until $J(z_{-i})<0$.

\begin{figure*}[t]
    \centering
    \includegraphics[width=0.9\textwidth]{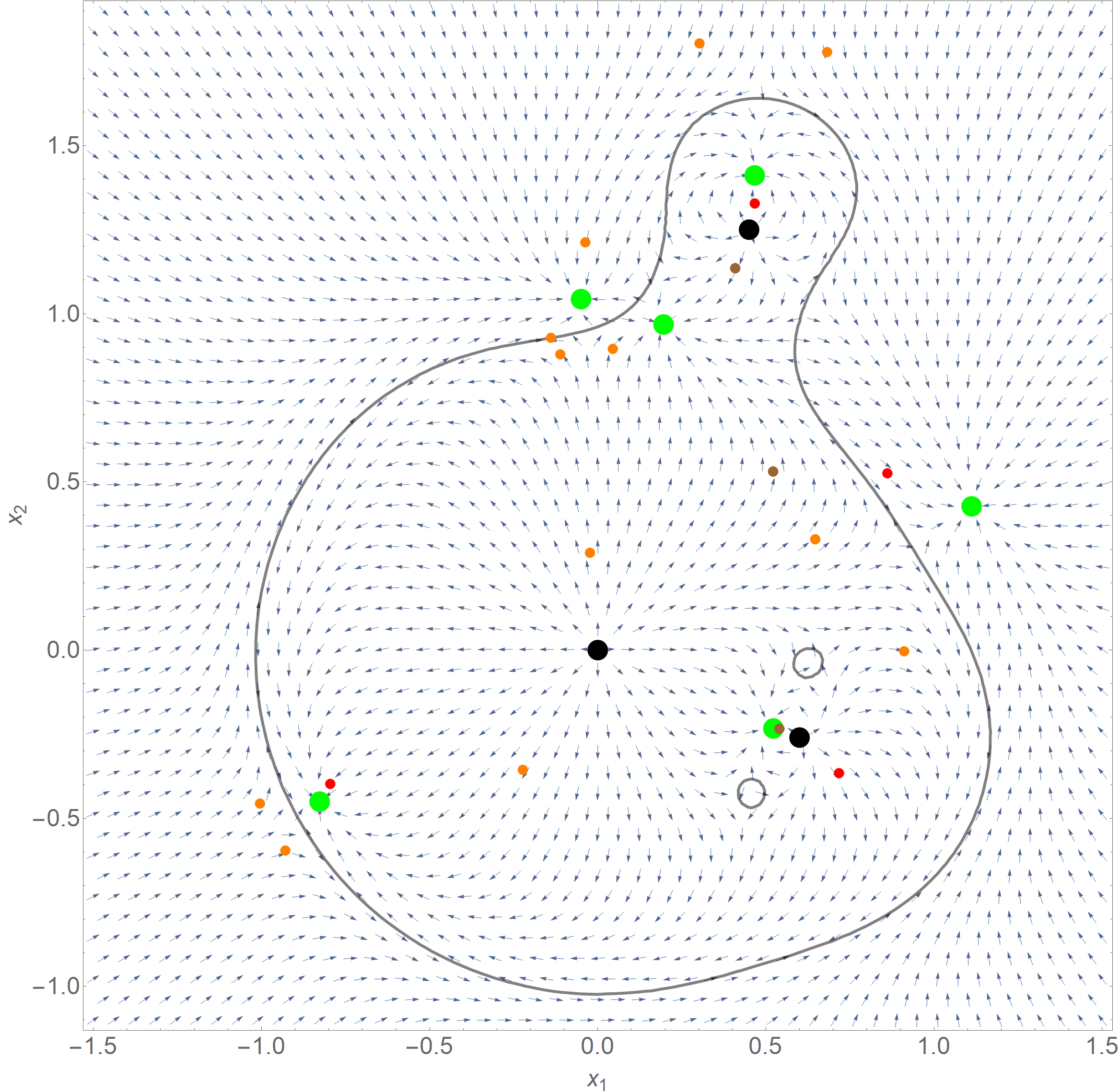}
    \caption{Solution search in the \texttt{Nopoly} algorithm. Here we have three lenses (black dots) with their critical curves in gray. The initial seeds based on image perturbations (Section \ref{Sec basic images}) are the red dots. The seeds for central images (Section \ref{Sec central images}) are the brown dots. The vectors indicate the direction of displacement $\epsilon$ as calculated by the Newtons-Raphson method (\ref{epsilon}). Following the vector flow from the initial seeds, we find the images (green dots). For each image, we also calculate new seeds (orange dots) for possible partner images beyond the critical curves (Section \ref{Sec partner images}).  We note that the positive parity image on the right, the negative parity image on top and the negative parity image on the left are very close to the corresponding red seeds and are therefore immediately found. The remaining two negative parity images are caught starting from the brown seeds. At this point the positive parity image on the left would still be missing, but we see that the search for partner images starting from the nearby negative parity image places two orange seeds very close to this final image, which is eventually recovered. }
    \label{fig:complete}
\end{figure*}
\subsubsection{Seeds for central images}\label{Sec central images}

The guesses for negative parity images do not work any more when two or more lenses are much closer than their Einstein radii. In this regime, we rather have a common negative parity image relative to the total mass and a central image very close to the barycenter. While the common negative parity image is easily recovered by whatever guess from Eq. (\ref{z-i}), we need additional initial seeds to make sure that the central images are not missed. Therefore, for each pair of lenses satisfying the criterion
\begin{equation}
    |a_i-a_j|<m_i+m_j,
\end{equation}
we calculate the guess
\begin{equation}
    z_{ij}=\frac{m_i a_j + m_j a_i}{m_i+m_j}.
\end{equation}
If $J(z_{ij})<0$, this is added to the list of initial conditions.

\subsubsection{Looking for partner images}\label{Sec partner images}

Here we come to the core of the strategy to find all images. Once we have the images found by running our modified Newton-Raphson algorithm on the first two sets of initial conditions discussed above, for each image we look for possible partners on the other side of the closest critical curve. Indeed, if the source enters a caustic, an additional pair of images is formed with opposite parities. If we have already found one of the two in the first search, we have good chances to find the other just by looking on the other side of the critical curve. 

Operationally, for each image $z$ found starting from the already discussed guesses, we calculate the distance to the closest critical curve $\delta_0$ by Eq. (\ref{delta0}), introduced in the discussion of the corrections to the Newton-Raphson algorithm. This estimate is very effective and reliable, as already discussed before. Therefore, we propose two additional seeds as
\begin{equation}
    z_{a\pm} = z + \delta_0 (2\pm i),
\end{equation}
where $i$ is the imaginary unit displacing the guess on either side perpendicular to $\delta_0$. If $z$ is very far from all critical curves, this guess may fail from crossing it. So, if the sign of the Jacobian is still the same, we just make a few attempts with different factors multiplying $\delta_0$, otherwise we give up for this $z$ and move to the next image.

For each seed in this new set of additional guesses, we repeat the Newton-Raphson algorithm and find either duplicates or new images. For each new image found, we look for partner images in the same way: we add two more guesses and run Newton-Raphson again, until all guesses have been checked and no new images are found. A practical example of a full search is shown in Fig. \ref{fig:complete}.

\subsubsection{Extreme fail-safe random search}

We find that this algorithm for branching new initial conditions from the images found from the first guesses is extremely effective in retrieving all images for whatever multiple lenses. The fact that we have at least an image in each critical curve thanks to the $z_{-i}$ set makes this algorithm very robust.

However, since there is no way to check that the search has been fully successful, we always check that the parities of the roots follow the theorem on the number of images, namely
\begin{equation}
    n_- - n_+ = N-1.
\end{equation}
The number of negative parity images minus the number of positive parity images must equal the number of lenses minus one.

If this fundamental theorem is not fulfilled, then we are missing at least one image. We have thus included an extreme fail-safe random search that looks around each lens with guesses:
\begin{equation}
    z_{i,k,\lambda} = a_i + \lambda \sqrt{m_i} e^{i k \pi/6},
\end{equation}
with $k=0,1, ..., 11$ and $\lambda$ starting from 1 with growing oscillations between values larger and smaller than one.

Of course, if we end in this pit of the \texttt{Nopoly} algorithm, there is no guarantee of the computational time required to find the missing image. Luckily, we find it to happen in exceptional cases with the final solutions making perfect sense anyway. This is the price to pay for an algorithm that is not based on a polynomial search with a definite number of roots.

\section{Comparison and performance}\label{Sec performance}

The computational efficacy of microlensing algorithms is coupled to both the multiplicity of gravitational lenses and the magnification regime. This section presents an evaluation of the performance metrics for the three distinct algorithms —\texttt{Singlepoly}, \texttt{Multipoly}, and \texttt{Nopoly}— with the objective of ascertaining the most efficient and reliable methodology across a diverse spectrum of lens configurations. Each algorithm exhibits a unique set of computational strengths and limitations, rendering them differently suited for modeling multi-lens gravitational systems.

\begin{figure*}[t]
    \centering
    \includegraphics[width=1\textwidth]{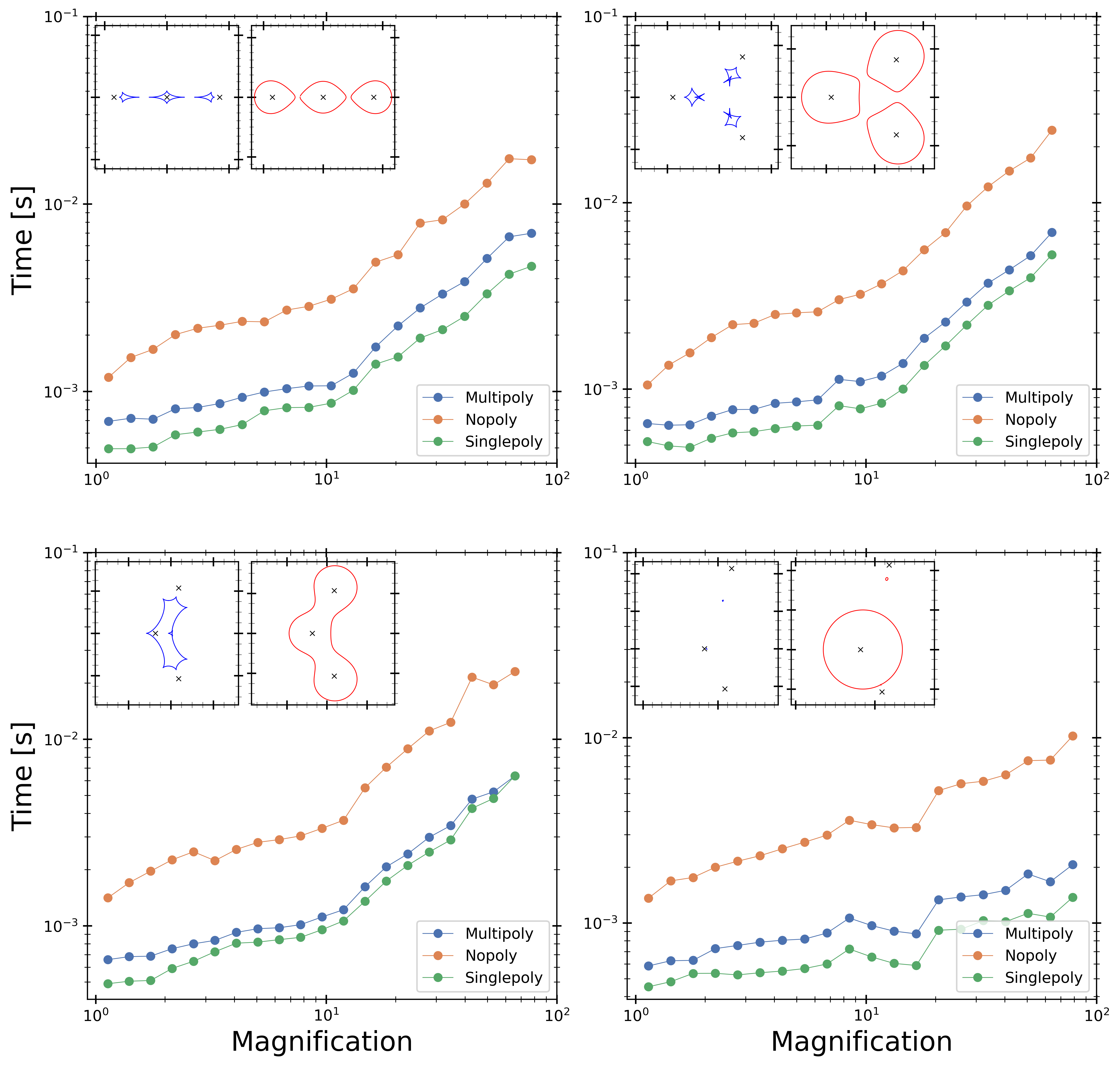}
    \caption{Comparisons of the performance of the \texttt{Singlepoly}, \texttt{Multipoly}, and \texttt{Nopoly} algorithms across different levels of magnification in a three-lens system. The points reflects the averaged computational performance across various source positions within the same magnification bin. Four different lens configurations were selected, characterized by the parameters ($q_2$,$q_3$,$s_2$,$s_3$,$\theta$,$\rho$). Here, $q_2$ and $q_3$ represent the mass ratios relative to the largest mass, $s_2$ and $s_3$ denote the distances of these masses from the largest mass, $\theta$ is the angle between the lines connecting the two smaller masses to the largest mass and $\rho$ is the source radius.The configurations are $A\left[1,1,1.7,1.7,\pi,10^{-3}\right]$, $B\left[1,1,1.5,1.5,\pi/3,10^{-3}\right]$,$C\left[1,1,1.2,1.2,0.7\pi,10^{-3}\right]$,$D\left[3.3*10^{-6},10^{-3},1,2,0.7\pi,10^{-3}\right]$.The images in the top left of each graph show the caustics and critical curves for each lens configuration. The computational performance of the three algorithms in a three-lens configuration remains consistent across different lens configurations. The order of efficiency, with \texttt{Singlepoly} being the fastest and \texttt{Nopoly} the slowest, is maintained regardless of the specific lens parameters.}
    \label{fig:times_comparisons}
\end{figure*}

Fig. \ref{fig:times_comparisons} provides a comprehensive visualization of the performance characteristics of the \texttt{Singlepoly}, \texttt{Multipoly}, and \texttt{Nopoly} algorithms as functions of the magnification in a triple lens system. The analysis encompasses four distinct lens geometries: a linear configuration, an equilateral triangular arrangement, an isosceles triangular disposition, and a planetary system analog. In the planetary system configuration, we simulate a primary lens (analogous to a stellar mass) accompanied by two secondary lenses with mass ratios approximating those of Earth and Jupiter relative to the Sun, respectively. The \texttt{Singlepoly} algorithm demonstrates superior computational velocity but exhibits, as previously mentioned, diminished reliability, particularly in scenarios characterized by high mass ratios where keeping numerical precision becomes critical. Conversely, the \texttt{Multipoly} algorithm manifests enhanced stability and precision, rendering it a more robust option, albeit at the cost of marginally increased computation time. The \texttt{Nopoly} algorithm, in the context of triple lens configurations, emerges as the least computationally efficient of the three, with a computation time about 5 times longer than the \texttt{Multipoly} algorithm.

\begin{figure}[t]
    \centering
    \includegraphics[width=0.5\textwidth]{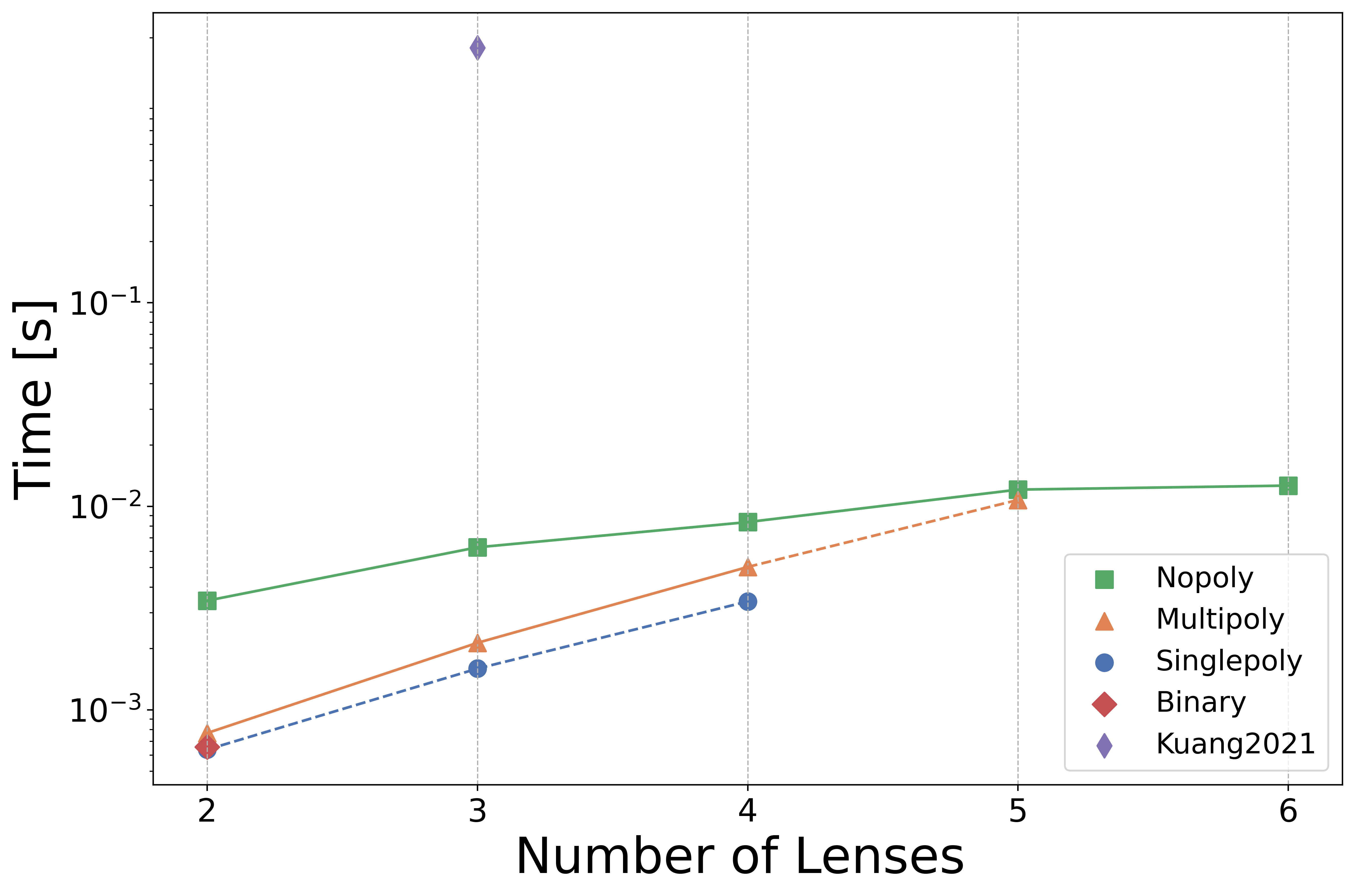}
    \caption{Comparison of the three algorithms-\texttt{Singlepoly}, \texttt{Multipoly}, and \texttt{Nopoly}- as the number of lenses in the system varies. The analysis includes \texttt{VBBinaryLensing} for binary lenses and the \texttt{triplelens} code published by \cite{Kuang2021}. Time estimates were based on calculations across all source positions for lens configurations producing a magnification of $20\pm5$. The computations were repeated for four different configurations, and the average times obtained were then used as data. In the case of binary lenses, \texttt{VBBinaryLensing} remains the most performant code. For triple lenses, \texttt{Singlepoly} is the most efficient but suffers from numerical precision issues. The dashed lines indicate that beyond that number of lenses, the algorithm's reliability decreases. Therefore, the preferred choice is \texttt{Multipoly}, which is slightly slower but more reliable. As the number of lenses increases \texttt{Multipoly} significantly degrades in performance, making \texttt{Nopoly} the reference code.}
    \label{fig:times_Nlens}
\end{figure}

As the lens multiplicity increases, the computational dynamics change significantly, as illustrated in Fig. \ref{fig:times_Nlens}. This analysis focuses on intermediate magnification events of order 20, with computation times averaged across four distinct lens geometries. In triple lens systems, the \texttt{Multipoly} algorithm maintains its efficiency, but its computational time grows substantially as the number of lenses increases. For systems with five or more lenses, the \texttt{Nopoly} algorithm becomes more competitive, ultimately surpassing the \texttt{Multipoly} algorithm in terms of computational efficiency, making it the preferred method for higher-order lens systems.

Furthermore, Fig. \ref{fig:times_Nlens} incorporates comparative analyses with pre-existing algorithms, including \texttt{VBBinaryLensing} for binary systems and the \texttt{triplelens} code published in \citet{Kuang2021}. In the case of binary lens systems, \texttt{VBBinaryLensing} retains its position as the optimal choice, having been specifically optimized for such configurations. In contrast, the triple lens code from \citet{Kuang2021} exhibits significantly inferior performance relative to the algorithms presented here, with computational times exceeding those of our methods by a factor of approximately $10^3$.

We caution the reader that the analysis presented in this section is to be intended to be informative of the status at the time of publication of this article. All codes will undergo further optimizations in the near future to improve their performance. We will keep track of these progresses on the GitHub repository in near real time.

\section{Conclusions}\label{Sec conclusions}
In this study, we have introduced and analyzed three innovative algorithms for multiple microlensing calculations: \texttt{Singlepoly}, \texttt{Multipoly}, and \texttt{Nopoly}. These approaches represent a significant leap forward in our ability to model and analyze complex microlensing events, offering unique and powerful solutions to the computational challenges that have long impeded progress in this field. Our comprehensive analysis across various lens configurations yields significant insights into their performance, applicability, and limitations. For binary lens systems, the established \texttt{VBBinaryLensing} remains the gold standard, outperforming all other algorithms due to its highly specialized optimization for this specific scenario. 

\texttt{Singlepoly} extends the traditional polynomial approach to multiple lenses, building upon established techniques in binary lens modeling. This algorithm demonstrates remarkable computational speed, making it an attractive option for rapid analysis of large datasets. In triple lens configurations, \texttt{Singlepoly} exhibits the highest computational efficiency. However, its reliability diminishes in scenarios involving high mass ratios. This limitation underscores the need for more robust methods in certain microlensing regimes, especially when dealing with complex lens configurations or extreme mass ratios that may be indicative of exotic planetary systems.

\texttt{Multipoly} addresses the limitations of \texttt{Singlepoly} by introducing a dynamic re-centering technique, designed to enhance accuracy in multi-lens calculations. By recalculating the polynomial in local coordinate systems centered on each lens, \texttt{Multipoly} significantly improves numerical stability, particularly in triple lens systems. This algorithm strikes a balance between computational efficiency and precision, offering a more reliable alternative to \texttt{Singlepoly} at a modest increase in computation time. As a result, \texttt{Multipoly} emerges as the preferred choice for triple and quadruple lens systems, providing a robust balance between accuracy and computational speed.

\texttt{Nopoly} represents a paradigm shift in microlensing calculations, employing a modified Newton-Raphson method directly on the lens equation without polynomial transformation. This novel approach circumvents many of the numerical issues associated with polynomial root-finding, offering a scalable solution for systems with higher lens multiplicity. While initially less efficient for triple and quadruple lens systems, \texttt{Nopoly}'s performance improves dramatically as the number of lenses increases. For systems with five or more lenses, its direct approach to the lens equation proves increasingly advantageous, ultimately outperforming polynomial-based methods like \texttt{Multipoly} in efficiency.

\texttt{VBMicrolensing} has already been incorporated in the modeling platform \texttt{RTModel}\footnote{\url{https://github.com/valboz/RTModel}} \citep{RTModel}, but we believe it will be useful to many more platforms for large-scale simulation and modeling. The development and implementation of these algorithms have significant implications for the field of microlensing. As we enter an era of increasingly precise photometric observations, exemplified by upcoming missions like the {\it Roman} Space Telescope, the ability to efficiently and accurately model complex lens systems becomes paramount. Our suite of algorithms provides researchers with a flexible toolkit to tackle a wide range of microlensing scenarios, from simple binary lenses to intricate multi-planetary systems. This enhanced computational capability could lead to new discoveries in exoplanetary science, potentially revealing subtle planetary signatures that were previously obscured by computational limitations.
Moreover, the scalability of the \texttt{Nopoly} algorithm opens up exciting possibilities for exploring extreme microlensing scenarios. Systems with multiple planets and exomoons can now be modeled with unprecedented efficiency, paving the way for more thorough and nuanced analyses of microlensing observations. As we continue to push the boundaries of exoplanet detection and characterization, these computational advancements will play a crucial role in unlocking the full potential of microlensing as a probe of planetary systems across our galaxy.

Finally, we remark that the \texttt{VBMicrolensing} project will continue its development with the addition of new features and optimizations following the requests from the microlensing scientific community. We invite the reader to check our public repository to stay tuned with the release of future versions.

\begin{acknowledgements}
We acknowledge financial support from PRIN2022 CUP D53D23002590006.
\end{acknowledgements}



\bibliographystyle{aa} 
\bibliography{VBMicrolensing.bbl}

\end{document}